\documentclass[12pt]{article}
\pdfoutput=1
\usepackage{Max}

\title{K3 metrics from little string theory}
\author[1]{Shamit Kachru}
\author[2]{Arnav Tripathy}
\author[1]{Max Zimet}
\affil[1]{Stanford Institute for Theoretical Physics,

Stanford University, Stanford, CA 94305 USA

~}
\affil[2]{Department of Mathematics,

Harvard University, Cambridge, MA 02138 USA}

\date{}

\newcommand{\ccwarrow}{\text{\Large$\curvearrowleft$}}

\begin{document}

\maketitle

\begin{abstract}
Certain six-dimensional (1,0) supersymmetric little string theories, when compactified on $T^3$, have moduli spaces of vacua given 
by smooth K3
surfaces.  Using ideas of Gaiotto-Moore-Neitzke, we show that this provides a systematic procedure for determining 
the Ricci-flat metric on 
a smooth K3 surface in terms of BPS degeneracies of (compactified) little string theories.

\end{abstract}

\newpage
\tableofcontents
\hypersetup{linkcolor=blue}

\section{Introduction}


In this paper, we discuss a new, physically motivated, systematic procedure for determining Ricci-flat metrics on smooth K3 surfaces.
Our approach builds upon a series of papers by Gaiotto, Moore, and Neitzke \cite{GMN:walls,GMN:classS,GMN:framed,GMN:2d,GMN:networks,GMN:snakes} (see also the review \cite{neitzke:hkReview}) devoted to improving our understanding of 4d $\N=2$ gauge theories and their BPS states. One of their main results is a relationship between these BPS states and the hyper-K\"ahler moduli space $\M$ of the theory compactified on a circle $S^1_R$ of radius $R$ (or, rather, the Coulomb branch thereof -- we henceforth neglect Higgs branches). This can be intuitively understood in the case where all BPS states are mutually local (i.e., all electrically charged in some duality frame).  In this case, BPS states of the 4d theory provide instantons of the theory on a circle that correct the na\"ive metric obtained by neglecting KK modes \cite{sw:3d,vafa:spacetimeInsts,seiberg:mirrorT}, which we call the semi-flat metric for reasons to be explained shortly. A non-renormalization theorem \cite{seiberg:mirrorT} guarantees that there are no further corrections. Of course, things are more complicated when there are mutually non-local BPS states, and this is where the ideas of Gaiotto, Moore, and Neitzke prove invaluable.  A proper accounting of all BPS states yields the full set of corrections at large $R$ to the semi-flat metric.

\medskip
Our main contribution is to replace the 4d $\N=2$ field theory with a 6d $\N=(1,0)$ little string theory \cite{s:hlst} (see also the reviews \cite{aharony:lst,kutasov:lst}) compactified on a 2-torus (and the theory compactified on a circle with the little string theory compactified on a 3-torus). In order to distinguish between the little string theory on $T^2$ and $T^3$, we will refer to the former as the 4d theory and the latter as the 3d theory, even though they both have six total dimensions. While it is the case that at low energies the 4d theory reduces to a 4d $\N=2$ gauge theory, adopting the latter viewpoint restricts us to an infinitesimal, non-compact piece of the total moduli space. In contrast, the Coulomb branch of the compactified little string theory is compact \cite{w:smallInst,s:fBranes,s:K3,intriligator:compact}. Indeed, the moduli space of the 3d theory (in the specific case of interest to us) is a K3 surface.

\medskip
In order to obtain the metric on the Coulomb branch of the 4d ${\cal N}=2$ theory on a circle, Gaiotto, Moore, and Neitzke construct special locally holomorphic functions on the moduli space and use wall crossing formulae to determine their discontinuities. The idea of building up the moduli space (without its metric) by studying rings of locally-defined holomorphic functions and the way that these rings glue together will hardly surprise mathematicians -- this is the standard construction of a ringed space. In particular, this approach has been used \cite{fukaya:insts,KS:gluing,GrossSiebert} to study the Strominger-Yau-Zaslow (SYZ) picture of mirror symmetry in the large complex structure limit \cite{strominger:mirrorT} (which for us is the $R \to \infty$ limit). For K3 surfaces, this approach was pioneered by \cite{KS:gluing}, where the symplectomorphisms that eventually became the wall crossing formulae of \cite{KS:walls,GMN:walls,GMN:framed} first appeared. We therefore regard this paper as completing a circle of ideas that has been nearly a decade and a half in the making.

\medskip
A parallel history concerns the K3 metric. The semi-flat metric appeared in \cite{greene:cosmicString}. As part of the SYZ conjecture, \cite{strominger:mirrorT} noted that corrections thereto away from the large complex structure limit should be determined by holomorphic disc instantons in the mirror manifold. As we will describe below, these instanton corrections map (via string dualities) to the instanton corrections to the compactified little string theory that we discussed above.  Such corrections by the lightest BPS states of the little string theory -- which yield a surprisingly accurate approximation to the true metric -- were studied (albeit not in this language) by \cite{gross:OV}. 

\medskip
The more complete study of disc instantons that govern the determination of the K3 metric has been under investigation by Lin \cite{lin:thesis,lin:walls1,lin:walls2,lin:walls3,lin:walls4,lin:walls5}. However, it has been speculated that these counts will grow too quickly to provide a convergent formula for a smooth, Ricci-flat K3 metric. Our main mathematical contribution is to provide a physical proof that this is not the case. Furthermore, as we discuss in the conclusion, by relating this problem to BPS state counting in little string theory, we open up new possibilities for the determination of these counts. Finally, we note that this paper constitutes a proof (at a physical level of rigor) of the SYZ conjecture for K3 surfaces at the level of the metric (as opposed to, say, only the complex structure).

\medskip
The rest of the paper is organized as follows. In section \ref{sec:LST}, we discuss our physical setup and its relationship with K3 and mirror symmetry. In section \ref{sec:wallMetric}, we review the results of \cite{GMN:walls} that allow us to determine the K3 metric (and point out that they are unmodified by the passage from 4d field theory to 6d compactified little string theory).\footnote{As we explain in more detail below, this method only provides the metric away from singular points of the semi-flat metric. However, physically it is clear that the metric can be extended over these points; this is worked out concretely for some examples in \cite{garza:singular1,garza:singular2}.} 
We then discuss approximations to the complete metric in section \ref{sec:approx} and compare our results with the philosophy of \cite{gross:OV}. We conclude in section \ref{sec:conclude} with ideas for future research.

\section{Little string theory and K3} \label{sec:LST}

The set of little string theories comprises a rich zoo, but we will be interested in one particular family.  
We will view the little string theory of interest as describing the worldvolume of an instanton 5-brane (i.e. a 5-brane associated to an instanton in the transverse dimensions) in the $SO(32)$ heterotic string theory at zero string coupling (but finite string scale $M_s$).  This gives us a number of ways to intuitively understand its relationship with K3. First, at low energies the 6d theory reduces to a $\N = (1,0)$ $Sp(1)\cong SU(2)$ gauge theory, in accordance with heterotic-type I duality which takes this 5-brane to a D5-brane \cite{w:smallInst}. In addition to a free hypermultiplet, which parametrizes the center of mass of the 5-brane, there are 16 fundamental hypermultiplets, corresponding to the 16 D9-branes of the type I theory.\footnote{One can also arrive at the number 16 by searching for a 6d $\N=(1,0)$ $SU(2)$ field theory without tensor multiplets and with a non-gravitational UV completion \cite{s:6d}.} Of course, since this theory is non-renormalizable it is only an effective field theory, but it provides a useful explanation for why the Coulomb branch $\M$ of the little string theory on $T^3$ is a K3 surface. Namely, $SU(2)$ Wilson lines around $T^3$ parametrize $T^3/Z_2$ (where the quotient accounts for the Weyl group), and dualizing the photon contributes an extra scalar which upgrades the moduli space to $T^4/Z_2$ \cite{s:K3,sw:3d}. Alternatively, T-duality allows us to replace the D5-brane with a D2-brane probing $T^3/Z_2$, and from the perspective of this brane the extra dimension provided by the dual photon corresponds to the extra circle of M-theory. Of course, this reasoning is semi-classical, and quantum corrections can desingularize the surface. Again, string theory makes this manifest, as the M-theory description of this setup is a K3 compactification with an M2-brane probe \cite{s:K3}. At generic points in moduli space, the K3 surface is smooth.

\medskip
String dualities also provide useful descriptions of the 4d theory obtained by taking $R\to \infty$. In this case, we have heterotic strings on $T^2$, and $SU(2)$ Wilson lines on $T^2$ parametrize a Coulomb branch $\B$ that is classically $T^2/Z_2$, which is topologically $S^2$. This $T^2/Z_2$ can again be probed after T-dualizing the D5-brane, which now becomes a D3-brane \cite{s:fBranes}. At low energies, the theory reduces to a non-linear sigma model to this moduli space, which is endowed with a K\"ahler structure by virtue of 4d $\N=2$ supersymmetry and so is properly regarded as $\PP^1$. However, away from singular points $\B^{\rm sing}\subset \B$ where BPS states become massless, we can usefully describe the low-energy behavior of the theory locally (that is, near some point in moduli space) by a $U(1)$ gauge theory \cite{sw:theory1,sw}. Around singular points, the coupling constant of this $U(1)$ gauge theory undergoes monodromies, and in order to obtain a weakly-coupled description one might have to perform an $SL(2,\ZZ)$ electric-magnetic duality transformation. Geometrically, this behavior of the coupling constant describes the complex structure of the fiber of an elliptic fibration over the $\PP^1$ moduli space; the total space of this fibration is an elliptically fibered K3 surface, but it is important to emphasize that the volume of the fiber has no physical meaning. The quantum-corrected behavior of the D3-brane, which describes both the metric on the moduli space and the complex structure of this elliptic fibration, is again accounted for by a K3 compactification -- now of F-theory \cite{sen:FOrientifolds,dasgupta:constant}. (See also the reviews \cite{dabholkar:orientifolds,morrison:compactDual}.) However, the metric on this K3 surface \cite{greene:cosmicString} is a bit singular -- specifically, it is ill-defined at 24 singular fibers (which are 7-branes in F-theory). Because the torus fibers are flat, this metric is called the `semi-flat' metric.

\medskip
Finally, we introduce a circle transverse to the D3-brane of radius $\hat R$. T-dualizing it yields M-theory on an elliptically fibered K3 surface (i.e., with a section, corresponding to the marked point of the elliptic fibers, whose homology class we identify with that of the base) \cite{vafa:F,morrison:triples} the area of whose fibers is inversely proportional to $\hat R$. The D3-brane becomes an M5-brane wrapping the elliptic fibers \cite{w:MGaugeSol}. Taking $\hat R\to\infty$ to recover the setup of interest\footnote{While this reasoning works in this case, we note that such reasoning with infinitesimal compactifications is not always so innocent \cite{mz:BPSdecay}.} causes the fibers to collapse to zero volume (which is consistent with the fact that the IR 4d theory only cares about the complex structure of the fibers).

\medskip
Alternatively, we could have taken the circle to be wrapped by the D3-brane. In this case, we would return to the probe M2-brane setup and take the elliptic fibers to have vanishing area. This defines a 4d theory, where the fourth dimension emerges from large stringy corrections.

\medskip
However, we should emphasize that the K3 surfaces in the M2 and M5 frames are not identical; in fact, they are \emph{mirror}.\footnote{This distinction might seem confusing, since if we remove the probe brane the passage from F-theory to M-theory is in both cases implemented by compactifying a circle, T-dualizing, and lifting from IIA to M-theory, and so the K3 surfaces in the two frames should be the same. The resolution of this conundrum is that for $\M$ a smooth elliptically fibered K3 surface, mirror symmetry coincides with hyper-K\"ahler rotation, which leaves the metric invariant \cite{bruzzo:hkRot,huybrechts:hkRot}. Once we include Wilson lines, the passage from F-theory to M-theory is not so straightforward and the distinction between $\M$ and its mirror becomes important.} This is clear from the SYZ picture of mirror symmetry as fiberwise T-duality \cite{vafa:k3inst,strominger:mirrorT}, since the M5-brane wraps the Seiberg-Witten curve\footnote{Generally speaking, the Seiberg-Witten curve $\Sigma$ is a non-compact Riemann surface with punctures that encode flavor information. The torus fibration over the Coulomb branch is then related to the Jacobian of its compactification $\bar\Sigma$ (or, more precisely, the Prym variety associated to the Hitchin fibration). The lattice of gauge charges is $H_1(\bar\Sigma,\ZZ)$, while the lattice of gauge and global charges is a subquotient of $H_1(\Sigma,\ZZ)$ \cite{GMN:classS}. While it would be interesting to find a $\Sigma$ whose first homology encodes the information about the singular fibers (analogous to the spectral curve of a Hitchin system), which presumably accounts for the effects of the singular fibers on the procedure that took us from the D3-brane to the M5-brane, here we will content ourselves with a discussion of the compact special Lagrangian torus $\bar\Sigma$. However, we note that the M5-brane frame suggests \cite{w:MGaugeSol} that the relevant generalization of the Hitchin system is that of \cite{donagi:megaHitchin}.\label{ft:spectral}} $\bar\Sigma$, whereas the D3- and M2-brane probes see its Jacobian $J(\bar\Sigma)$. Indeed, if we compactify a circle transverse to the M2-brane and wrapped by the M5-brane and take this to be the M-theory circle, then we find T-dual D2- and D4-branes. The moduli space $\M$ of our compactified little string theory whose metric we study is the K3 surface probed by the M2-brane. We call the mirror $\M^\vee$.

\medskip
Besides allowing us to intuit the role played by K3 in the low energy behavior of the compactified little string theory, these duality frames also let us relate BPS states in the little string theory to the geometry of K3.\footnote{One might wonder if we need to consider excited states of branes in order to match the `oscillator spectrum' of the little string theory, so that the geometry of a brane does not suffice to characterize its BPS states. (The quotes remind us that these are not the usual oscillator states of perturbative string theory.) That this is not the case follows from the fact that exciting a brane costs energy, and so for an excited state to be BPS it needs a new conserved charge, but we have geometric pictures for all of the conserved charges. So, the Hagedorn degeneracy does not preclude us from being able to label states by their geometrically-interpretable conserved charges. Finally, we note that geometry reproducing dual oscillator behavior is familiar in related contexts, e.g. \cite{yauZaslow}.} In the M5-brane frame, BPS states correspond to M2-branes ending on the M5-brane and wrapping holomorphic 2-cycles with one leg along the base and one along the fiber \cite{yi:surfaces,mikhailov:surface,sethi:FWebs}. These 2-cycles can terminate at singular fibers. In the D3-frame, these BPS states correspond to string webs along the Coulomb branch \cite{bergman:FWebs,sethi:FWebs,zwiebach:webs}. (The same geometric picture holds for many BPS states in the M2-brane case, except now this M2-brane will merge with the M2-brane giving the BPS state \cite{sw:3d,sethi:FWebs}.)

\medskip
After compactifying on $S^1_R$ in the M5-brane frame, the M2-brane BPS states give rise to worldsheet instantons ending on a D4-brane. These should be counted by open string reduced Gromov-Witten invariants $\Omega(\gamma)$ labeled by their gauge and global (or flavor) charges\footnote{We restrict to non-singular points in the parameter space of the little string theory (i.e., the moduli space of its parent heterotic string theory), so that the K3 surface is smooth. This means that our flavor symmetries are abelian.} $\gamma=\gamma^g\oplus \gamma^f \in H_1(\bar\Sigma,\ZZ)\oplus H_2(\M^\vee,\ZZ)/\avg{b,f} = H_2(\M^\vee, \bar\Sigma; \ZZ)/\avg{b}$, where $b,f$ are respectively the base and fiber classes.\footnote{Technically, this direct sum is a bit sloppy, as the splitting of $H_2(\M^\vee,\bar\Sigma;\ZZ)$ is not canonical and does not hold globally, in the sense that monodromies in the Coulomb branch will mix up gauge and flavor charges, as described in \S\ref{sec:data}.} Invariants which presumably are these $\Omega(\gamma)$ have recently been defined and studied by Lin \cite{lin:walls1,lin:walls2,lin:walls3,lin:walls4,lin:walls5}. In particular, he shows that they may be defined so that they satisfy the wall crossing formula of \cite{KS:walls}, which plays a central role in the techniques of Gaiotto, Moore, and Neitzke. We thus see that our construction of the metric concretely realizes the proposal of \cite{strominger:mirrorT,fukaya:insts,KS:gluing,GrossSiebert} that holomorphic disc instantons in the mirror determine corrections to the semi-flat metric of $\M$,\footnote{The BPS states may not have the topology of a disc in spacetime \cite{yi:surfaces,mikhailov:surface,sethi:FWebs,zwiebach:webs}. `Disc' here refers to the topology of the worldsheet associated to the Gromov-Witten invariants $\Omega$.} which has been verified in special cases, including toric Calabi-Yau manifolds \cite{auroux:insts1,auroux:insts2,chan:discs,chan:discs2}.

\medskip
Before proceeding, we mention a few more interesting aspects of the physics of little string theory. First, everything we have said generalizes trivially if we have $r$ heterotic 5-branes instead of one. For example, the 6d gauge theory has gauge group $Sp(r)$. The other duality frames similarly have $r$ branes instead of one, and the moduli space is now the symmetric product\footnote{The quotient by the symmetric group $S_r$ arises in the 6d field theory language because $S_r$ is a subgroup of the Weyl group of $Sp(r)$.} $\Sym^r(K3)$ \cite{aharony:questions,douglas:moreD3s}. The low energy abelian gauge theory that locally describes the physics of the 4d Coulomb branch is now a $U(1)^r$ gauge theory, the Coulomb branch is $2r$-dimensional, the fibers are now $2r$-tori, and the total space has a semi-flat metric which is corrected by 4d BPS states upon compactification on a circle. While the main object of interest in this paper is the K3 metric, we emphasize that the arguments of this paper apply equally well to $\Sym^r(K3)$.

\medskip
Second, the growth of the density of states as a function of energy is characterized by Hagedorn behavior. For BPS states, whose mass equals the absolute value of their central charge $Z_\gamma$, this implies that the growth of $\Omega$ is at most
\be \log |\Omega(\gamma)| \sim \frac{\sqrt{r} |Z_\gamma|}{M_s} \ . \label{eq:hagedorn} \ee
(The combination $r/M_s^2$ is the 't Hooft coupling of the 6d gauge theory.) This expression plays an essential role in this paper, as there is an important condition for the convergence of certain expressions in \cite{GMN:walls} that is roughly
\be \log |\Omega(\gamma)| < 2\pi R |Z_\gamma| \label{eq:condition} \ ,\ee
which in light of \eqref{eq:hagedorn} reduces, in our case, to
\be R > \frac{\sqrt{r}}{2\pi M_s} \label{eq:lstCondition} \ .\ee
More precisely, an equivalent condition to that of \cite{GMN:walls} is \cite{stoppa:DT,bridgeland:RH}
\be \sum_{\gamma\in \hat\Gamma} |\Omega(\gamma)| e^{-2\pi R|Z_\gamma|} < \infty \ , \label{eq:actualCondition} \ee
and this follows from \eqref{eq:hagedorn} and \eqref{eq:lstCondition}. We thus see that our results should only apply in a neighborhood of the large complex structure locus, where the fibers are small. It is physically unsurprising that there is an important change when $R\sim \sqrt{r}/2\pi M_s$, since little string theories have T-duality. In contrast, the 4d $\N=2$ field theories have $M_s=\infty$. However, this suggests a powerful approach not available for the field theory near $R=0$: we can study the $R\to \infty$ limit of a T-dual little string theory! Finally, we note that \eqref{eq:actualCondition} is not expected to hold in gravitational theories, as the analogue of \eqref{eq:hagedorn} would have charge \emph{squared} on the right hand side. As stressed in \cite{neitzke:hkReview}, this poses an important obstacle to the study of stringy quaternionic-K\"ahler moduli spaces. This obstacle has been attributed \cite{pioline:inst} to neglected (KK and NS) 5-brane instantons, whose contributions to the metric are in general unknown (but see \cite{pioline:nsInst,alexandrov:nsInsts,pioline:ns5wall} for progress on this problem). Little string theory's Hagedorn growth is just slow enough for the ideas of \cite{GMN:walls} to apply.

\section{Wall crossing and the metric} \label{sec:wallMetric}

We now briefly summarize the formalism of \cite{GMN:walls}\footnote{See \cite{pioline:linearHK,pioline:linearQK,pioline:dInst,alexandrov:dInst,pioline:qkhk,alexandrov:metric} for related results.} that determines the metric at large $R$ and then assure the reader that nothing changes when we upgrade to little string theory. For the benefit of mathematicians (and logical clarity), we mimic \cite{neitzke:hkReview} and separate the input data and the output results into separate subsections.

\subsection{Input} \label{sec:data}

Consider a rank $r$ 4d $\N=2$ gauge theory. The Coulomb branch $\B$ is an $r$-complex-dimensional K\"ahler manifold with a $T^{2r}$-fibration over it. The latter has singular fibers at a complex codimension 1 locus $\B^{\rm sing}\subset \B$ where BPS states become massless, and we define $\B' = \B \backslash \B^{\rm sing}$. We denote by $\M$ the total space of this fibration, and by $\M' \subset \M$ the total space of the fibration over $\B'$. We will use $u$ to denote a point in $\B'$. At such a point $u$, the low energy behavior of the theory is captured by a pure $U(1)^r$ gauge theory whose coupling constants are related to the complex structure of the $T^{2r}$ fiber over $u$. Technically, these fibers are not merely tori, but abelian varieties -- that is, they are K\"ahler manifolds whose K\"ahler forms have integral periods; physically, the latter come from the integral symplectic pairing on the charge lattice whose existence is guaranteed by the Dirac quantization condition. However, as observed in \cite{sw:3d}, when we compactify on $S^1_R$ there are instanton corrections to the K\"ahler structure of $\M$, and in addition there is a whole twistor sphere's worth of complex structures, most of which also receive instanton corrections (as we discuss below). So, for our purposes $\B$ is a complex manifold (which we will still endow with some extra structure, using the central charge) with a $T^{2r}$-fibration $\pi : \M' \to \B'$ over it.

\medskip
We now discuss the gauge, global, and combined charge lattices\footnote{While we call them lattices, physics provides no natural inner product in general, and so really we only have the structure of finitely-generated free abelian groups.\label{ft:noIP}} in more detail. Thanks to monodromies around $\B^{\rm sing}$, these lattices are not constant over $\B'$, but are instead fibers of local systems -- i.e., fibrations over $\B'$ with monodromies around $\B^{\rm sing}$ determined by a flat connection (which for a vector bundle is the same as a representation of the fundamental group of $\B'$). We will write expressions such as $\gamma\in \Gamma$ in order to indicate that $\gamma$ is a local section of the local system $\Gamma$. The local systems of gauge, global, and combined charges, which we denote $\Gamma,$ $\Gamma_{\rm flavor}$, and $\hat\Gamma$, fit into an exact sequence
\be 0 \to \Gamma_{\rm flavor} \to \hat\Gamma \to \Gamma \to 0 \ .\ee
However, $\Gamma_{\rm flavor}$ is a trivial local system (i.e. the fibers are all the same lattice and there is no monodromy). A local section of $\hat\Gamma$ splits (non-canonically) as a local section of $\Gamma \oplus \Gamma_{\rm flavor}$, but it need not be the case that $\hat\Gamma = \Gamma \oplus \Gamma_{\rm flavor}$ globally -- monodromies not only act nontrivially on gauge charges, but they can add linear combinations of gauge charges to global charges, since a combined gauge and global symmetry is physically the same as just the global symmetry \cite{sw}. A fiber over the point $u\in \B$ is of the form $\hat\Gamma_u \cong \Gamma_u \oplus (\Gamma_{\rm flavor})_u$, where $\Gamma_u \cong \ZZ^{2r}$ is the lattice of electromagnetic charges\footnote{Actually, it would be more natural to take $\Gamma_u$ to be the lattice of electromagnetic charges labeling IR Wilson-'t Hooft line defects, since we will shortly associate line operators to elements of $\Gamma_u$ \cite{GMN:framed}. Equivalently, this is the lattice of electromagnetic charges in the presence of arbitrary BPS line defects of the UV theory at the origin of space (since from far away a charge orbiting a line defect simply looks like a combined line defect). The existence of such a lattice follows from the mutual locality condition on line operators. This contains the genuine electromagnetic charge lattice $\tilde\Gamma_u$ as a sublattice (since we allow there to be no UV line defect), to which the symplectic pairing has a restriction, and is itself contained in $\tilde\Gamma_u\otimes \RR$. Of course, $\Omega(\gamma;u)=0$ if the projection of $\gamma\in \hat\Gamma_u$ to $\Gamma_u$ is not contained in $\tilde\Gamma_u$.

\medskip
The need for this larger lattice can be appreciated by considering the example of a pure $SU(2)$ gauge theory. All BPS states (such as the W-boson) have even electric charge, but BPS states in the presence of a Wilson line defect in the fundamental representation carry odd electric charge \cite{GMN:framed}.

\medskip
Anyways, since considering the line operators associated to elements of $\tilde\Gamma_u$ will suffice, we henceforth neglect the distinction between $\Gamma_u$ and $\tilde\Gamma_u$ and refer to $\Gamma_u$ as the gauge charge lattice.} and $(\Gamma_{\rm flavor})_u$ is the lattice of flavor charges. $\hat\Gamma_u$ has an integer-valued anti-symmetric pairing $\avg{,}$, and $(\Gamma_{\rm flavor})_u$ is its radical -- i.e., the lattice of charges $\hat\gamma$ for which $\avg{\hat\gamma, \gamma}=0$ for all $\gamma\in \hat\Gamma_u$ -- since these charges have no magnetic duals.

\medskip
The next data physics provides is a central charge function, which is a global section $Z\in \Hom(\hat\Gamma,\CC)$ varying holomorphically over $\B'$. A local section $\gamma\in \hat\Gamma$ defines a local holomorphic function $Z_\gamma(u) = Z(u) \cdot \gamma$, where $Z(u)\in \hat\Gamma_u^* \otimes \CC$. The latter has a natural interpretation as the direct sum of a period vector in $\Gamma_u^*\otimes \CC$ and a vector of mass parameters in $(\Gamma_{\rm flavor})^*_u\otimes\CC$. The mass parameters are actually $u$-independent. Mathematically, the central charge function induces an $S^1$-worth of singular symplectic integral affine structures on $\B$. That is, for each $\varphi\in \RR/2\pi\ZZ$ we can use $\Real(e^{i\varphi}Z_{\gamma^g})$ for $\gamma^g\in \Gamma$ as coordinates for $\B$, and the transition functions \cite{sw}
\be \Real(e^{i\varphi} Z_{\gamma^g}) \to \Real(e^{i\varphi} Z_{N \gamma^g + M \gamma^f}) = \Real(e^{i\varphi} Z_{N\gamma^g}) + \Real(e^{i\varphi} Z_{M\gamma^f}) \ , \ee
where $N\in \Sp(2r,\ZZ)$ and $M\in \Hom(\Gamma_{\rm flavor},\Gamma)$, are elements of $\Sp(2r,\ZZ) \ltimes \ZZ^{2r}$. For the sake of completeness, we note that the central charge function satisfies a number of other conditions; see, e.g., \cite{neitzke:hkReview}.

\medskip
Finally, we require the BPS index, which is a function $\Omega: \hat\Gamma \to \ZZ$. (More precisely, this is the second helicity supertrace. It is a signed count of short multiplets, which in particular increases by 1 for each massive hypermultiplet and decreases by 2 for each massive vector multiplet.) CPT implies $\Omega(\gamma; u) = \Omega(-\gamma; u)$. Hagedorn growth -- or anything slower -- implies \eqref{eq:actualCondition}, as we have already discussed.

\medskip
For later convenience, we introduce the M\"obius transform $\bar\Omega : \hat\Gamma \to \QQ$ of $\Omega$, defined by
\be \bar\Omega(\gamma;u) = \sum_{n\ge 1, \, n|\gamma} \frac{1}{n^2}\Omega\parens{\frac{\gamma}{n};u} \ ,\quad 
\Omega(\gamma; u) = \sum_{n\ge 1, n|\gamma} \frac{\mu(n)}{n^2} \bar\Omega\parens{\frac{\gamma}{n}; u} \ .\ee
Here, $\mu$ is the M\"obius function. The new invariants account for the universal $1/n^2$ factor associated in open Gromov-Witten theory to multi-coverings for disc instantons \cite{vafa:openGV,vafa:openGV2}. In addition, they have simpler wall crossing behavior than $\Omega$ \cite{sen:rationalDT}.

\medskip
All of this data must be consistent with the wall crossing formula \cite{KS:walls,GMN:walls} described below, which in particular is strong enough to determine $\Omega(\cdot; u)$ for all $u\in\B'$ given $\Omega(\cdot; u_0)$ for any $u_0\in \B'$. We define the set
\be \hat\Gamma'_u := \{\gamma\in \hat\Gamma_u : \Omega(\gamma;u)\not=0 \} \ee
of \emph{active charges at $u$} and
\be \ell_\gamma(u) := \{\zeta \in \CC^\times \  |\  Z_\gamma(u)/\zeta \in \RR_-\} \ ,\quad \gamma \in \hat\Gamma'_u \ . \label{eq:ell} \ee
Then, $\Omega(\gamma; u)$ is locally constant and can jump only at
\be W := \{ u\in\B' \ |\  \exists \gamma,\gamma' \in \hat\Gamma'_u : \ \avg{\gamma,\gamma'}\not=0 \, , \  Z_\gamma(u)/Z_{\gamma'}(u) \in \RR_+ \} \ , \ee
which is a union of real-codimension-1 loci in $\B'$, called walls of marginal stability, where there exist mutually non-local charges $\gamma,\gamma'\in \hat\Gamma'_u$ such that $\ell_\gamma(u) = \ell_{\gamma'}(u)$.

\medskip
In addition, the BPS index must satisfy the following ``support property" from \cite{KS:walls}, which is so named because it constrains the charges at which $\Omega$ can be supported. Namely, for any $u\in\B'$ and any (or equivalently, some) choice of positive-definite norm on $\hat\Gamma_u\otimes\RR$, there exists a $K>0$ such that
\be \|\gamma\| < K |Z_\gamma(u)| \ ,\quad \forall \gamma\in \hat\Gamma'_u \ . \label{eq:support} \ee
This can be motivated geometrically \cite{KS:walls}, and indeed is satisfied by holomorphic discs \cite{lin:walls2}, since holomorphic cycles are calibrated.

\medskip
We emphasize that all of the data discussed thus far pertains to the 4d theory -- we were focused on 4d charges, BPS states, etc.\footnote{The wall crossing formula is stated using symplectomorphisms $\K_\gamma$ of a space $T$ to be defined shortly. Functions $\X_\gamma$ appear in the definition \eqref{eq:K} of $\K_\gamma$. Since the $\X_\gamma$ depend on how we compactify -- i.e., they depend on $R$ and the parameters $m_{\gamma^f}$ to be introduced shortly -- it might seem that the wall crossing formula is not purely a statement about 4d physics. However, this is misleading. For the purposes of the wall crossing formula, one should think of the $\X_\gamma$ as elements of an abstract algebra satisfying \eqref{eq:commRing}, rather than as the concrete functions we will define below. This perspective makes it clear that the wall crossing formula only involves 4d physics.} We now compactify the 4d theory on a circle of radius $R$, and upon doing so we can introduce holonomies for background gauge fields. These parameters are specified by a homomorphism $\theta : \Gamma_{\rm flavor} \to \RR/2\pi \ZZ$. So, we have $\rank \Gamma_{\rm flavor}$ new mass parameters: $\theta_{\gamma^f} = 2\pi R m_{\gamma^f}$, where $\gamma^f\in \Gamma_{\rm flavor}$.

\subsection{Output}

The construction of the metric relies essentially on the fact that $\M$ is a hyper-K\"ahler manifold (which is guaranteed by 3d $\N=4$ supersymmetry). We briefly review some essential consequences of this and refer the reader to \cite{hitchin:hkSUSY} for additional background. $\M$ has a $\PP^1$-worth of K\"ahler structures, which we parametrize as
\be \omega^{(\zeta)} = \sum_{\alpha=1}^3 c_\alpha^{(\zeta)} \omega_\alpha \ ,\quad J^{(\zeta)} = \sum_{\alpha=1}^3 c_\alpha^{(\zeta)} J_\alpha \ ,\ee
\be c_\alpha^{(\zeta)} = \frac{1}{1+|\zeta|^2}\parens{ 2\Imag\zeta, -2\Real\zeta, 1-|\zeta|^2} \ ,\quad \sum_\alpha \parens{c_\alpha^{(\zeta)}}^2 = 1 \ .\ee
Here, $J_\alpha = (J_1, J_2, J_3)$ are a triplet of complex structures that satisfy the quaternion algebra
\be J_\alpha J_\beta = \epsilon_{\alpha\beta\gamma} J_\gamma - \delta_{\alpha\beta} \ , \label{eq:quat} \ee
and $\omega_\alpha = (\omega_1, \omega_2, \omega_3)$ are compatible K\"ahler forms in the sense that
\be \omega_\alpha(v_1, v_2) = g(J_\alpha v_1, v_2) \ , \quad g(v_1,v_2) = \omega_\alpha(v_1, J_\alpha v_2) \ ,\quad J_\alpha = - g^{-1}\omega_\alpha \ , \label{eq:triple} \ee
where $g$ is the metric (which is the same for all $\zeta$) and $v_1,v_2$ are tangent vectors to $\M$. If we want to refer to $\M$ as a complex manifold with a fixed complex structure specified by $\zeta$, we will use the notation $\M_\zeta$. 

\medskip
A useful notion is the twistor space $\Z$, which is the total space of the holomorphic fibration $p:\Z\to \PP^1$ with fiber $\M_\zeta$. As a manifold, this is simply the product $\M\times S^2$. Finally, for $\zeta\in \CC^\times$ (i.e., $\zeta\not=0,\infty$, to which we generally restrict) it is also useful to re-package the K\"ahler forms as
\be \varpi(\zeta) = -\frac{i}{2\zeta} \omega_+ + \omega_3 -\frac{i}{2} \zeta \omega_- \ , \label{eq:holoSymp} \ee
where
\be \omega_\pm = \omega_1 \pm i\omega_2 \ . \label{eq:sympIJ} \ee
$\varpi(\zeta)$ is a holomorphic symplectic form on $\M_\zeta$ and defines a holomorphic section of the line bundle $\Omega_{\Z/\PP^1}^2\otimes p^*\Oo(2)$ over $\Z$. (The reason for the $\Oo(2)$ factor is that for $\zeta=0,\infty$ we have to rescale $\varpi$ respectively by $\zeta$ or $1/\zeta$.) We can clearly obtain the K\"ahler forms from \eqref{eq:holoSymp}; \eqref{eq:quat} allows us to also obtain the complex structures. For example, $J_\alpha^2=-1$ (no sum on $\alpha$) implies $J_\alpha=\omega_\alpha^{-1}g$, and so $J_3=J_1 J_2=-\omega_1^{-1}\omega_2$. So, once we have \eqref{eq:holoSymp}, \eqref{eq:triple} gives us the metric:
\be g = \omega_3 J_3 = -\omega_3 \omega_1^{-1} \omega_2 \ . \label{eq:metric} \ee
Determining \eqref{eq:holoSymp} will therefore be our goal. We also note that the antipodal map on $\PP^1$, $\zeta\mapsto -1/\bar\zeta$, lifts to an anti-holomorphic involution of $\Z$ that fixes the fibers pointwise, and $\nu^*\varpi = \overline\varpi$. $\nu$ is called the real structure of $\Z$, as it takes the complex structure on $\Z$ to its conjugate, and so it can be used to identify `real' objects as those which are $\nu$-invariant.

\medskip
We now introduce coordinates on the $T^{2r}$ fibers of $\M$ that are adapted for describing the semi-flat metric. Locally, we can choose a basis of sections $\{\gamma^i\}$, $i=1,\ldots,2r$, for $\Gamma$ compatible with a Lagrangian decomposition $\Gamma \cong \Gamma^m\oplus \Gamma^e$ into magnetic and electric charges. That is, $\epsilon^{ij} := \avg{\gamma^i,\gamma^j}$ is given by
\be \epsilon^{ij} = \twoMatrix{0}{{\bf 1}}{-{{\bf 1}}}{0} \ ; \ee
we denote the negative of the inverse matrix by $\epsilon_{ij} = \epsilon^{ij}$. We parametrize a position in the fiber over $u$ using $2r$ angles $\theta_{\gamma^i}\in \RR/2\pi \ZZ$. We then extend this definition to a twisted unitary character, i.e. a map $\theta : \Gamma_u \to \RR/2\pi\ZZ$ satisfying
\be e^{i(\theta_\gamma + \theta_{\gamma'})} = (-1)^{\avg{\gamma,\gamma'}} e^{i\theta_{\gamma + \gamma'}} \ .\ee
The space of such maps is topologically $T^{2r}$. Gluing together these spaces gives a local system of $2r$-tori which we identify as $\pi : \M' \to \B'$. Locally choosing a lifting of $\Gamma$ to $\hat\Gamma$, we can now use our homomorphism $\theta : \Gamma_{\rm flavor} \to \RR/2\pi \ZZ$ to extend $\theta$ to a twisted unitary character of $\hat\Gamma_u$.

\medskip
We would like to identify the moduli $\theta_\gamma$, where $\gamma\in \Gamma$, with electric and magnetic Wilson lines around $S^1_R$ of the IR 4d abelian gauge theories. However, the latter are more naturally thought of as untwisted unitary characters, i.e. homomorphisms $\tilde\theta : \Gamma_u \to \RR/2\pi\ZZ$. We would therefore like to identify the bundle of twisted unitary characters of $\hat\Gamma$ with the local system $\hat\Gamma^* \otimes (\RR/2\pi \ZZ)$. While this need not be possible globally, we can at least do so locally on $\B'$ by introducing a quadratic refinement $\sigma : \hat\Gamma_u \to \ZZ_2$ of the pairing $(-1)^{\avg{\gamma,\gamma'}}$, i.e. a map obeying
\be \sigma(\gamma)\sigma(\gamma') = (-1)^{\avg{\gamma,\gamma'}} \sigma(\gamma + \gamma') \ .\ee
In fact, it is necessary to do so in order to define the path integral of a gauge theory in this self-dual formalism where we simultaneously work with both electric and magnetic Wilson lines \cite{w:m5,freed:selfDual,singer:selfDual,moore:selfDual}. A natural way to obtain a quadratic refinement is to choose a local duality frame, $\hat\Gamma\cong \Gamma^m \oplus\Gamma^e \oplus \Gamma_{\rm flavor}$, decompose local sections accordingly as $\gamma = \gamma^m + \gamma^e + \gamma^f$, and define
\be \sigma(\gamma) = (-1)^{\avg{\gamma^e, \gamma^m}} \ . \label{eq:canonicalRefine} \ee
In any case, having made a choice of $\sigma$, we may non-canonically map a unitary character to a twisted unitary character using \cite{GMN:walls,GMN:framed}
\be e^{i\theta_\gamma} = e^{i\tilde\theta_\gamma}\sigma(\gamma) \ . \label{eq:thetaRel} \ee
In some cases, such as those discussed in \cite{GMN:classS,GMN:framed}, there exists a global choice of $\sigma$, but this need not necessarily always be the case.\footnote{However, \cite{Chuang:2013wt,sen:noExotics} have made progress in proving the related `no exotics' conjecture \cite{GMN:framed}. It might be valuable to see if this extends to little string theory.} Anyways, this is the last we will have to say about quadratic refinements -- we will henceforth simply work with twisted unitary characters.

\medskip
In order to define convenient coordinates on $\M_\zeta$, we now consider Wilson-'t Hooft line operators in the IR abelian gauge theory. In general, the expression for a supersymmetric Wilson line in a representation $\R$ is \cite{nekrasov:CSdim,maldacena:wLoops,rey:wLoops,kapustin:holoLoops}
\be \Tr_\R \, P \exp \oint \parens{ \frac{\varphi}{2\zeta} + A + \frac{\zeta \bar\varphi}{2}} \ ,\ee
where $\varphi$ is a vector multiplet scalar in the complexified adjoint representation and $A$ is a gauge field (which we take to be anti-Hermitian so that it defines a covariant derivative $d+A$). This can be thought of as describing the insertion of an infinitely massive particle in the representation $\R$ with central charge of phase\footnote{Physically, we should impose $|\zeta|=1$. We then analytically continue the functions $\X_\gamma(\zeta)$ to be defined shortly to all of $\CC^\times$.} $\zeta$. It is $1/2$-BPS, and the preserved supercharges depend on $\zeta$. By wrapping the analogous Wilson-'t Hooft line $L^{\rm IR}_{\gamma,\zeta}$ associated to an infinitely massive dyon of charge $\gamma = \gamma^g \oplus \gamma^f \in \Gamma_u\oplus \Gamma_{\rm flavor}$ around $S^1_R$ and taking the expectation value, we obtain functions
\be \avg{L^{\rm IR}_{\gamma,\zeta}} = \X_\gamma(u,\theta;\zeta) = \exp \brackets{ \frac{\pi R}{\zeta} Z_{\gamma^f} + i \theta_{\gamma^f} + \pi R \zeta \overline{Z_{\gamma^f}}} \X_{\gamma^g}(u,\theta;\zeta) \ . \label{eq:LIRx} \ee
For later convenience, we also introduce the notation
\be \Y_\gamma = \log \X_\gamma \ . \label{eq:Y} \ee
In the $R\to \infty$ limit, we obtain the semi-flat coordinates
\be \X^{\rm sf}_\gamma = \exp \brackets{ \frac{\pi R}{\zeta} Z_\gamma + i \theta_\gamma + \pi R \zeta \overline{Z_\gamma}} \ , \quad \gamma\in\hat\Gamma \ . \ee
For, as the length of the Wilson loop tends to infinity, quantum fluctuations become irrelevant and we can treat the dynamical fields appearing in the Wilson loop as if they were background fields. Another way to explain this physically is by regarding $S^1_R$ as the time direction, so that we can identify $\X_\gamma(u,\theta;\zeta)$ as the following trace \cite{GMN:framed}:\footnote{The presence of $\theta$, rather than $\tilde\theta$, in this expression concretely demonstrates our earlier statement that the path integral depends on a choice of quadratic refinement, thanks to \eqref{eq:thetaRel}.}
\be \Tr_{\Hh_{u;\gamma,\zeta}} (-1)^F e^{-2\pi R H} e^{i\theta_\Q} \ .\ee
Here, $\Q$ is the charge operator, $H$ is the Hamiltonian, $(-1)^F$ imposes supersymmetric boundary conditions (i.e. fermions are periodic, like bosons) on the path integral, and $\Hh_{u;\gamma,\zeta}$ is the Hilbert space of the theory defined in the presence of $L^{\rm IR}_{\gamma,\zeta}$. Then, at large $R$ this projects onto the ground state, which satisfies a modified BPS bound \cite{GMN:framed}
\be -2\pi R H = \frac{\pi R}{\zeta}Z_\gamma + \pi R \zeta \overline{Z_\gamma} \ ,\ee
thanks to the presence of the line operator, and we reduce to the case where we compute the worldline action of a particle in a classical background.

\medskip
These line operator expectation values allow us to determine the metric. We first provide the formulae that determine the metric, and then we explain the properties of the line operators that make this construction work. Recalling our local basis of sections for $\Gamma$, we obtain the following coordinates on $\M$:
\be \Y^i(\zeta) = \Y_{\gamma^i}(\zeta) \ .\ee
They are Darboux (or canonical) coordinates, as
\be \varpi(\zeta) = \frac{1}{8\pi^2 R} \epsilon_{ij} d'\Y^i(\zeta) \wedge d'\Y^j(\zeta) \ , \label{eq:darboux} \ee
where $d'$ denotes a differential that treats $\zeta$ as a constant, provides a holomorphic sympletic form on $\M_\zeta$. (To see what makes these coordinates canonical, compare this expression with the familiar symplectic form $dq^i\wedge dp_i$.) This induces the Poisson bracket
\be \{\X_\gamma,\X_{\gamma'}\} = 4\pi^2R \,\epsilon^{ij}\, \frac{\partial \X_\gamma}{\partial \Y^i}\frac{\partial\X_{\gamma'}}{\partial \Y^j} = 4\pi^2R \avg{\gamma,\gamma'} \X_\gamma \X_{\gamma'} \ ,\quad \gamma,\gamma'\in \hat\Gamma \label{eq:poisson} \ee
on $\M'_\zeta$. Note that $\varpi(\zeta)$ is independent of our splitting of $\hat\Gamma$ into $\Gamma\oplus\Gamma_{\rm flavor}$ -- that is, if $\gamma^i$ and $\tilde\gamma^i$ differ by an element of $\Gamma_{\rm flavor}$, then it follows from \eqref{eq:LIRx} that $d'\Y_{\gamma^i}(\zeta)=d'\Y_{\tilde\gamma^i}(\zeta)$.

\medskip
Before continuing, we would like to explain a limitation of this method. As we have mentioned, the expressions we provide below converge only for large enough $R$. As discussed in \cite{neitzke:hkReview}, `large enough' might depend on $u$. But, of course a fixed metric has the same $R$ for all $u$. So, practically one should fix an $R$ and then employ our results on the complement of the union of small open sets surrounding the points of $\B^{\rm sing}$. The ideas of \cite{garza:singular1,garza:singular2} should allow one to extend the metric over all of $\M$. Physically, the reason for this limitation is that the scale at which the 4d IR abelian gauge theory is trustworthy depends on $u$, and certain arguments (e.g. \eqref{eq:UVIR}) require this scale to be larger than $1/R$.

\medskip
The functions $\X_\gamma(\zeta)$ are mathematically useful for a number of physical reasons. First, let us be a bit more careful about the sense in which they are functions. Consider the set $T_u$ of maps $X : \Gamma_u \to \CC^\times$ satisfying
\be X_\gamma X_{\gamma'} = (-1)^{\avg{\gamma,\gamma'}} X_{\gamma + \gamma'} \ ,\ee
called the space of twisted complex characters of $\Gamma_u$. Topologically, $T_u$ is a complex algebraic torus $(\CC^\times)^{2r}$ parametrized by $X_{\gamma^i}$. These may be pieced together to form a local system of complex tori over $\B'$. In particular, if $\gamma$ and $\gamma'$ are local sections respectively defined on open subsets $U$ and $V$ of $\B'$ which agree on $U\cap V$, then $X_\gamma$ and $X_{\gamma'}$ agree on $U\cap V$. We let $T$ be the pullback via $\pi$ of this local system to $\M'$. (That is, local sections of $T$ can depend on $\theta$.) $\X(\zeta)$ are then local sections of $T$, meaning that they satisfy\footnote{In \cite{GMN:walls}, $\X$ is defined so that the twist is absent from this equation. However, the definition of $\X$ used in this paper is more canonical -- physically, because it is with our definition that $\X_\gamma$ is the expectation value of a line operator, and mathematically because it is independent of a choice of quadratic refinement \cite{GMN:framed}.}
\be \X_\gamma \X_{\gamma'} = (-1)^{\avg{\gamma, \gamma'}} \X_{\gamma + \gamma'} \ . \label{eq:commRing} \ee
Physically, the twist expresses the fact that the fermion number of a bound state of particles with charges $\gamma,\gamma'$ is shifted by $\avg{\gamma,\gamma'}$ \cite{GMN:walls}. If $\gamma$ is a local section defined on some open set $U\subset \B'$, then $\X_\gamma(\zeta)$ is holomorphic (actually, piecewise holomorphic, as discussed below) over $\pi^{-1}(U)$ in complex structure $\zeta$. Physically, this follows from the fact that $L^{\rm IR}_{\gamma,\zeta}$ is BPS \cite{GMN:framed}.

\medskip
Next, we note that
 \be \X_\gamma(\zeta) = \overline{\X_{-\gamma}(-1/\bar\zeta)} \ .\ee
This can be understood physically as arising from CPT. Mathematically, it is a reality condition chosen so that $\nu$ is anti-holomorphic and $\nu^*\varpi = \overline\varpi$.

\medskip
Physically, the reason that the $\X_\gamma$ have a natural Poisson structure can be understood from a non-commutative deformation of the algebra \eqref{eq:commRing},
\be \X_\gamma \X_{\gamma'} = y^{\avg{\gamma, \gamma'}} \X_{\gamma+\gamma'} \ , \label{eq:nonComm} \ee
which is obtained by including a chemical potential in the expectation value defining $\X_\gamma$ that keeps track of the grading by a certain conserved charge \cite{gukov:refineMotive1,gukov:refineMotive2,vafa:topWalls,GMN:framed}. For, the leading correction to \eqref{eq:nonComm} away from $y=\pm 1$ is given by $\braces{\X_\gamma, \X_{\gamma'}}$. That is, \eqref{eq:nonComm} coincides with the deformation quantization of the algebra of holomorphic functions on $\M'_\zeta$ associated to the Poisson bracket \eqref{eq:poisson}.\footnote{Deformation quantization involves the promotion of functions to non-commutative Heisenberg picture operators, which are allowed to be formal power series in a variable $\hbar$ (which in our case is given by $y=e^{i\pi\hbar}$).}

\medskip
The most important property of these functions $\X_\gamma$ is that they are only piecewise holomorphic, but their discontinuities are given by symplectomorphisms so that the holomorphic symplectic form $\varpi(\zeta)$ is smooth. More precisely, they can be discontinuous at the locus
\be L := \{(u,\zeta)\in \B'\times\CC^\times \ | \ \exists \gamma\in \hat\Gamma'_u : \zeta\in \ell_\gamma(u)\} \ .\ee
This can be understood by studying wall crossing for \emph{framed BPS states} \cite{GMN:framed}, i.e. BPS states which exist in the presence of a line operator. For, UV line operators flow in the IR to a sum of line operators, and the coefficients in this expansion are the BPS index counting framed BPS states with charge $\gamma$:
\be L^{\rm UV}_\zeta \mapsto \sum_{\gamma\in\hat\Gamma_u} {\overline{\underline{\Omega}}}(L^{\rm UV}_\zeta, \gamma, u) L^{\rm IR}_{\gamma,\zeta} \ . \label{eq:UVIR} \ee
When these line operators are wrapped around $S^1_R$ we have
\be \avg{L^{\rm UV}_\zeta} = \sum_{\gamma\in\hat\Gamma_u} {\overline{\underline{\Omega}}}(L^{\rm UV}_\zeta, \gamma, u) \X_\gamma(u,\theta;\zeta) \ . \ee
Framed wall crossing occurs at the rays $\ell_{\gamma}(u)$. But, $\avg{L^{\rm UV}_\zeta}$ does not exhibit wall crossing, since the UV theory does not undergo a phase transition. So, $\X_\gamma$ must have discontinuities that cancel the effects of framed wall crossing \cite{GMN:framed}.

\medskip
In fact, an intuitive physical picture (based on an analogous gravitational story \cite{denef:walls}) determines both the location (i.e., $\ell_{\gamma}(u)$) and the form of these discontinuities \cite{GMN:framed}. We imagine a framed BPS state as being comprised of an infinitely massive core orbited by a halo of vanilla (non-framed) mutually BPS particles. A field theoretic version of the attractor equation yields the orbit radii of the halo's constituents, and this classical picture is trustworthy when these radii are large. Fortunately, framed wall crossing occurs in this regime -- specifically, at a ray $\ell_{\gamma}$, the radius for all particles whose charge is a positive multiple of $\gamma$ reaches infinity, and so all framed BPS states whose halos contain such a particle decay. Quantizing these halo configurations yields the framed wall crossing formula, and thus the discontinuities in $\X$.

\medskip
In order to state the framed wall crossing formula, we introduce the following birational symplectomorphisms $\K_\gamma : T \to T$ for $\gamma\in \hat\Gamma$:
\be \K_\gamma := \X_{\gamma'} \mapsto \X_{\gamma'} (1 - \X_\gamma)^{\avg{\gamma',\gamma}} \ . \label{eq:K} \ee
We can also describe $\K_\gamma$ in another way. First, to each $\gamma\in \hat\Gamma$ we associate an infinitesimal symplectomorphism generated by the Hamiltonian $-\X_\gamma/4\pi^2R$, i.e. by
\be e_\gamma := f \mapsto \frac{1}{4\pi^2R}\braces{\X_\gamma, f} \ .\ee
By the Jacobi identity, this satisfies
\be \brackets{e_\gamma, e_{\gamma'}} = (-1)^{\avg{\gamma,\gamma'}} \avg{\gamma, \gamma'} e_{\gamma+\gamma'} \ . \label{eq:bracket} \ee
That is, the $e_\gamma$ generate a Lie algebra that satisfies \eqref{eq:bracket}. We can then identify
\be \K_\gamma = \exp \sum_{n=1}^\infty \frac{1}{n^2} e_{n\gamma} \ .\ee
This is useful in deriving a number of consequences of the wall crossing formula \cite{GMN:walls}. It also highlights the important property
\be \avg{\gamma,\gamma'} = 0 \quad \Rightarrow \quad [\K_\gamma, \K_{\gamma'}] = 0 \ . \label{eq:commute}\ee
The framed wall crossing formula is then the statement that when $u\not\in W$ the discontinuity of $\X$ at a ray $\ell$ is given by
\be \X^+ = S_\ell(u) \, \X^- \ ,\quad S_\ell(u) := \prod_{\gamma \in \hat\Gamma_u' : \, \ell_{\gamma}\!(u) = \ell } \K_{\gamma}^{\Omega(\gamma;u)} \ . \label{eq:framedJumps} \ee
Here, $\X^+,\X^-$ are the limits of $\X$ as $\zeta$ approaches $\ell$ clockwise or counterclockwise, respectively. Thanks to \eqref{eq:commute}, since $u\not\in W$, there is no ordering ambiguity. $S_\ell=1$ for all but a countable set of rays, since $\hat\Gamma_u$ is countable.

\medskip
The (vanilla) wall crossing formula can now be intuitively phrased as the requirement that $\X$ be discontinuous only at $L$, and in particular not at $(W\times\CC^\times) \backslash L$. That is, all discontinuities of $\X$ are as in \eqref{eq:framedJumps}. At a point $u\in W$, multiple rays $\ell_\gamma(u)$, comprising a set $\ell(u)$, coincide at a single ray, so we should be more careful with the definition of $S_\ell(u)$. We therefore deform slightly away from $u$ to $u'\not\in W$, define $\ell(u')$ to be the set
\be \ell(u') = \{ \  \ell \  | \  \exists \gamma\in \hat\Gamma'_u : \ell_\gamma(u) \in \ell(u) \,,\  \ell = \ell_\gamma(u') \} \ee
of deformed rays, and compute the monodromy as we cross all of these rays in a counterclockwise order (i.e., the product from right to left is in counterclockwise order):\footnote{See the discussions in \cite{KS:walls,GMN:walls,GMN:classS} for an explanation of how to carefully make sense of this product using \eqref{eq:support}.}
\be A_\ell(u') := \prod_{\tilde\ell \in \ell(u')}^\ccwarrow S_{\tilde \ell}(u') = \prod^\ccwarrow_{\substack{\gamma\in \hat\Gamma'_u : \\ \ell_\gamma(u)\in \ell(u)}} \K_\gamma^{\Omega(\gamma;u')} \ .\ee
The wall crossing formula is the statement that the symplectomorphism $S_\ell(u)$ obtained in the limit $u'\to u$ is the same regardless of on which side of the wall $u'$ resides. This is non-trivial, as the order of the rays in $\ell(u')$ is reversed as we cross the wall. To motivate this result, note that
\be \avg{L_\zeta^{\rm UV}}_u = S(P) \avg{L_{\zeta'}^{\rm UV}}_{u'} \ , \label{eq:trans} \ee
where $S(P)$ is a composition of symplectomorphisms given by the above framed wall crossing formula, and $P$ denotes a path in $\widehat{\B'}\times \widehat{\CC^\times}$ from $(u',\zeta')$ to $(u,\zeta)$. (The hats indicate that we are considering the universal covers of $\B'$ and $\CC^\times$. This is necessary for $\B'$ because of the monodromies we have discussed, and is necessary for $\CC^\times$ because an anomaly translates the Witten effect for line operators \cite{kapustin:lines,henningson:lines,s:lines} into monodromy of $L_\zeta^{\rm UV}$.) We want the framed wall crossing formula to imply that $S(P)$ depends only on the endpoints of $P$. This follows if the UV theory has sufficiently many line operators so that transformations of the form \eqref{eq:trans} determine $S(P)$; in the well-studied theories, this appears to be the case \cite{GMN:framed,cordova:lines}.

\medskip
Another important observation is that a non-renormalization theorem protects the $\zeta=0$ complex structure -- which may be identified with that of the 4d gauge theory's torus fibration -- from instanton corrections \cite{sw:3d}. So, for any $\gamma\in \hat\Gamma_u$ the asymptotics of $\X_\gamma$ are such that $\lim_{\zeta\to 0} \X_\gamma(\zeta) / \X_\gamma^{\rm sf}(\zeta)$ exists and is real. This implies that the holomorphic 2-form in complex structure $\zeta=0$ may be identified with that of the semi-flat limit after a fiberwise diffeomorphism \cite{GMN:walls}.

\medskip
These properties of $\X(\zeta)$, plus the semi-flat asymptotics 
\be \X_\gamma = \X_\gamma^{\rm sf}(1 + \Oo(e^{-const\cdot R})) \label{eq:sfAsymp} \ee
in regions bounded away from $\B^{\rm sing}$, turn out to completely determine the functions \cite{garza:singular2}. They also imply that \eqref{eq:darboux} gives a holomorphic symplectic form on $\M'_\zeta$. (For example, \eqref{eq:sfAsymp} implies that for large enough $R$, the holomorphic symplectic form is non-degenerate -- that is, its kernel is the $2r$-complex-dimensional subspace of the $4r$-complex-dimensional $T_\CC \M'_\zeta$ consisting of anti-holomorphic tangent vectors.) In order to determine $\X(\zeta)$ from these constraints, \cite{GMN:walls} showed that these constraints determine a family of Riemann-Hilbert problems on $\PP^1$ parametrized by the moduli $u\in \B'$ and parameters of the compactified little string theory (including $R$), which we collectively denote by $\Pp$, whose solution is the $\X_\gamma$. They then showed that the solution to this Riemann-Hilbert problem satisfies the following integral equation:\footnote{In passing from \eqref{eq:TBA} to \eqref{eq:TBA2}, we are forced to define $\ell_{\gamma'}(u)$ for $\gamma'\in \hat\Gamma_u \backslash \hat\Gamma'_u$. We can do so as in \eqref{eq:ell}. Equivalently, we can notice that for each $\gamma'$ that contributes to \eqref{eq:TBA2} there exists an $n\ge 1$ such that $n|\gamma'$ and $\gamma'/n\in \hat\Gamma'_u$; we can then define $\ell_{\gamma'}(u) = \ell_{\gamma'/n}(u)$, since the latter expression is independent of $n$.}\textsuperscript{,}\footnote{This is a form of the Thermodynamic Bethe Ansatz \cite{zamolodchikov:TBA}. The relationship between the TBA and instanton corrections to moduli spaces was investigated in \cite{alexandrov:TBA}.}\textsuperscript{,}\footnote{It is shown in \S5.6 of \cite{GMN:walls} that this equation allows one to obtain the first few terms in the Laurent series of $\log\X_\gamma$ about $\zeta=0$. In tandem with \eqref{eq:holoSymp} and \eqref{eq:darboux}, this yields explicit expressions for the triplet $\omega_\alpha$ of symplectic forms in terms of $\X_\gamma$.}
\begin{align}
\X_\gamma(\zeta) &= \X_\gamma^{\rm sf}(\zeta) \exp\brackets{- \frac{1}{4\pi i} \sum_{\gamma'\in \hat\Gamma'_u} \Omega(\gamma';u) \avg{\gamma,\gamma'} \int_{\ell_{\gamma'}(u)} \frac{d\zeta'}{\zeta'} \frac{\zeta' + \zeta}{\zeta' - \zeta} \log\parens{1 - \X_{\gamma'}(\zeta')}} \label{eq:TBA} \\
&= \X_\gamma^{\rm sf}(\zeta) \exp\brackets{\frac{1}{4\pi i}\sum_{\gamma'\in \hat\Gamma_u} \bar\Omega(\gamma'; u) \avg{\gamma, \gamma'} \int_{\ell_{\gamma'}(u)} \frac{d\zeta'}{\zeta'} \frac{\zeta'+\zeta}{\zeta'-\zeta} \,\X_{\gamma'}(\zeta')} \label{eq:TBA2} \ .
\end{align}
It is shown in \cite{GMN:walls}, using the Banach contraction principle, and making use of the conditions \eqref{eq:actualCondition} and \eqref{eq:support}, that for sufficiently large $R$ this integral equation has a unique solution which may be obtained by iteration.\footnote{Similar ideas appear in \cite{vafa:classification}.} That is, one starts with $\X^{(0)}_\gamma(\zeta) = \X_\gamma^{\rm sf}(\zeta)$, and then computes
\be \X_\gamma^{(\nu+1)}(\zeta) = \X_\gamma^{\rm sf}(\zeta) \exp\brackets{- \frac{1}{4\pi i}\sum_{\gamma'\in \hat\Gamma'_u} \Omega(\gamma'; u) \avg{\gamma, \gamma'} \int_{\ell_{\gamma'}(u)} \frac{d\zeta'}{\zeta'} \frac{\zeta'+\zeta}{\zeta'-\zeta} \log\parens{1 - \X_{\gamma'}^{(\nu)}(\zeta')}} \ .\ee
Furthermore, \cite{GMN:walls} provides a closed-form solution of \eqref{eq:TBA} in terms of a (possibly asymptotic, rather than convergent) sum over decorated rooted trees $\T$ -- that is, trees with a node singled out as the root, and where each node $a$ is decorated with a choice of $\gamma_a\in \hat\Gamma_u$. We denote an edge between nodes $a$ and $b$, where $a$ is closer to the root, as $(a,b)$, and the decoration of the root node as $\gamma_\T$. Then, the solution of \eqref{eq:TBA} is
\be \X_\gamma(\zeta) = \X_\gamma^{\rm sf}(\zeta) \exp \brackets{\sum_{\T} \G_\T(\zeta) \avg{\gamma, \W_\T}} \ , \label{eq:closed} \ee
where
\be \W_\T = \gamma_\T \, \frac{\bar\Omega(\gamma_\T;u)}{|{\rm Aut}(\T)|} \prod_{(a,b)\in {\rm Edges}(\T)} \bar\Omega(\gamma_b;u) \avg{\gamma_a,\gamma_b} \ee
and $\G_\T(\zeta)$ is determined inductively by
\be \G_\T(\zeta) = \frac{1}{4\pi i} \int_{\ell_{\gamma_\T}(u)} \frac{d\zeta'}{\zeta'} \frac{\zeta'+\zeta}{\zeta'-\zeta} \X^{\rm sf}_{\gamma_\T}(\zeta')\prod_A \G_{\T_A}(\zeta') \ , \ee
where $\T_A$ are the subtrees of $\T$ obtained by deleting the root, and where $\G_\T(\zeta)$ with $\T$ a tree with only one node is given by the same formula with the product over $A$ omitted. Plugging \eqref{eq:closed} into \eqref{eq:darboux} gives
\begin{align}
\varpi(\zeta) &= \varpi^{\rm sf}(\zeta) + \frac{1}{4\pi^2R}\epsilon_{ij}d'\Y^{{\rm sf, \, }i}(\zeta) \wedge \sum_\T \avg{\gamma^j, \W_T} d'\G_\T(\zeta) \nonumber \\
&+ \frac{1}{8\pi^2R} \sum_{\T,\T'} \avg{\W_T,\W_{T'}} d'\G_\T(\zeta)\wedge d'\G_{\T'}(\zeta) \ ,
\end{align}
where, as in \eqref{eq:darboux}, the differentials $d'$ treat $\zeta$ as a constant. In particular, the first approximation (associated to $\X^{(1)}_\gamma(\zeta)$), obtained by summing only over trees with one node, is
\begin{align}
\varpi^{(1)}(\zeta) &= \varpi^{\rm sf}(\zeta) + \frac{1}{16\pi^3 iR}\epsilon_{ij} d'\Y^{{\rm sf, \, }i}(\zeta) \wedge \sum_{\gamma\in \hat\Gamma'_u} \Omega(\gamma;u) \avg{\gamma^j, \gamma} \int_{\ell_\gamma(u)} \frac{d\zeta'}{\zeta'}\frac{\zeta'+\zeta}{\zeta'-\zeta} \frac{d'\X^{\rm sf}_\gamma(\zeta')}{1 - \X^{\rm sf}_\gamma(\zeta')} \nonumber\\
&- \frac{1}{128\pi^4R} \sum_{\gamma,\gamma'\in \hat\Gamma'_u} \Omega(\gamma;u)\Omega(\gamma';u) \avg{\gamma,\gamma'} \nonumber \\
&\qquad\qquad \int_{\ell_\gamma(u)} \frac{d\zeta'}{\zeta'}\frac{\zeta'+\zeta}{\zeta'-\zeta} \, \int_{\ell_{\gamma'}(u)} \frac{d\zeta''}{\zeta''}\frac{\zeta''+\zeta}{\zeta''-\zeta} \frac{d'\X^{\rm sf}_\gamma(\zeta')}{1 - \X^{\rm sf}_\gamma(\zeta')} \wedge \frac{d'\X^{\rm sf}_{\gamma'}(\zeta'')}{1 - \X^{\rm sf}_{\gamma'}(\zeta'')}\ . \label{eq:firstApprox}
\end{align}
(Of course, $d'$ treats $\zeta'$ and $\zeta''$, as well as $\zeta$, as constants.)

\medskip
We conclude this section by mentioning another proof of the wall crossing formula, provided in \cite{GMN:walls}. This uses the observation that physically-motivated differential equations satisfied by the $\X_\gamma$ (related to holomorphy and anomalous Ward identities) define a flat connection with singularities for the infinite-dimensional bundle of real-analytic functions on the torus fibers of $\M$ over $\PP^1\times \Pp$ of which the $\X_\gamma$ are covariantly constant sections. Equivalently, they determine an `isomonodromic family' of flat connections with singularities over $\PP^1$, which are related by parallel transport along $\Pp$. But, these connections determine the `generalized monodromy' or Stokes data that define the discontinuities of a Riemann-Hilbert problem. (See \cite{w:wild} for a discussion of Stokes phenomena and isomonodromic deformation.) The large $R$ limit of these connections is known explicitly, and parallel transporting the large $R$ Stokes data shows that the Stokes data for all $R$ is indeed that of the Riemann-Hilbert problem described above. Isomonodromic deformation of the connection over $\PP^1$ as we vary $u$ then implies the wall crossing formula.

\subsection{Upgrade to little string theory}

None of the above changes when we upgrade to little string theory.\footnote{We note three caveats to this statement. First, the little string theory of interest has particles (0-branes), strings (1-branes), and monopoles (2-branes), and correspondingly has UV line operators, surface operators, and 3-manifold operators, all of which can be wrapped on $S^1_R$, with any extra dimensions wrapping a cycle of $T^2$. All of these decompose into sums of IR line operators wrapping $S^1_R$ in the 4d effective field theory.

Second, globally-defined holomorphic functions do not exist on compact manifolds, and so UV defect operator expectation values must exhibit some interesting behavior. For example, they could themselves have discontinuities (at different locations from those of the IR line operators, in general). Given that this issue arises specifically for little string theories, but not Coulomb branches of (compactified) asymptotically free or conformal field theories, as the latter are non-compact, we suspect that this may be related to the fact that the definition of operators in a little string theory requires smearing in position space \cite{polchinski:UVIR,aharony:lstThermo,kapustin:lstNonLocal}. It would be quite interesting to investigate this further. For now, our lack of knowledge regarding defect operators in little string theories does not concern us, since we can consider defect operators in 6d effective field theories that are valid at intermediate energy scales (such as an $SU(2)$ gauge theory with 16 fundamental hypermultiplets, in some regions of parameter space), and in addition there are many other arguments for the wall crossing formula, such as the one from \cite{GMN:walls} outlined above, as well as those of \cite{sen:rationalDT,vafa:topWalls}. We note that the same reasoning as above implies that the defect operators in these 6d effective field theories cannot have globally holomorphic expectation values; this might reasonably have been expected, given our experience with line operators in the 4d infrared effective field theories.

Third, certain real mass parameters affect the semi-flat limit by generalizing it from an elliptic fibration to a genus one fibration. This is discussed in more detail in \cite{mz:K3HK}. Here, we simply note that the formulae defining the semi-flat hyper-K\"ahler structure do not change; all that changes is that monodromies can induce translations on fiber coordinates. Therefore, in \S\ref{sec:semiFlat} we focus on the elliptically fibered case.} In particular, the proof of the wall crossing formula in \cite{GMN:framed} works here without modification. This stands in contrast to the 4d $\N=2$ supergravity context, where an analogous proof of the wall crossing formula holds, but only after one replaces line operators by supermassive black holes \cite{moore:halo}. This is because in supergravity, one cannot neglect gravitational backreaction of heavy external particles that define a Wilson-'t Hooft line operator. The proof of \cite{GMN:walls} should also be adaptable to little string theory.

\medskip
Much of the data discussed in \S\ref{sec:data} is actually quite familiar to both string theorists and mathematicians. Recall that the moduli space of semi-flat metrics on elliptically fibered K3 surfaces is
\be \brackets{ O(\Gamma^{18,2})\backslash O(18,2) / (O(18)\times O(2)) } \times \RR_+ \times \RR_+ \ , \label{eq:ellipticMod} \ee
where $\Gamma^{p,q}$ refers to the even unimodular lattice with signature $(p,q)$. The last two factors respectively parametrize the size of the base and the size of the fibers. The moduli space of F-theory on such a K3 surface is \eqref{eq:ellipticMod} except for the last factor, since the fibers are not part of spacetime. This is dually the moduli space of heterotic string theory on $T^2$ (where $\RR_+$ in this frame corresponds to the coupling constant).\footnote{The map between F-theory and heterotic moduli has been worked out explicitly for both 10-dimensional heterotic theories on $T^2$ without Wilson lines (where the dual F-theory compactifications involve K3 surfaces with Shioda-Inose structures) \cite{cardoso:fMap,lerche:fMap,morgan:fTheory,sethi:fluxes}.} Since the moduli space of heterotic string theory (except for the coupling constant) becomes the parameter space of the little string theory, \eqref{eq:ellipticMod} (minus the last two factors) becomes the mass parameters in the central charge function. Similarly, the flavor charge lattice is the spacetime gauge charge lattice $\Gamma^{18,2}$.\footnote{As discussed in footnote \ref{ft:noIP}, the inner product on the charge lattice should be forgotten. We nevertheless denote it as $\Gamma^{18,2}$ in order to emphasize its connections with the spacetime gauge charge lattice and $H_2(\M^\vee,\ZZ)$.} (The gauge charge lattice is also quite simple, as $r=1$.) However, note that there are only 18 flavor central charges, not 20. This follows from the fact that $O(18,2)/(O(18)\times O(2))$ has complex dimension 18. So, the flavor central charge is not an arbitrary homomorphism from $\Gamma_{\rm flavor}$ to $\CC$.

\medskip
Similarly, the moduli space of Ricci-flat metrics on K3 surfaces is
\be \brackets{ O(\Gamma^{19,3})\backslash O(19,3) / (O(19)\times O(3)) } \times \RR_+ \ 	, \label{eq:k3Mod} \ee
where the final factor corresponds to the total volume, and by heterotic/M-theory duality this is also the moduli space of heterotic on $T^3$. Again neglecting the $\RR_+$ factor, we find that the little string theory gains 21 real parameters upon compactification on $S^1_R$. $\rank\Gamma^{18,2}=20$ of them correspond to Wilson lines for background gauge fields, and the 21st is $M_sR$. The condition of having only abelian flavor symmetries in the little string theory translates to having only abelian gauge symmetries in the parent heterotic string theory and to having a K3 surface without ADE singularities in the M-theory frame.

\medskip
Finally, as we emphasized above, much of the data (such as the central charge and the charge local systems) has nothing to do with the compactification to three dimensions. So, except for the BPS index (which does not affect the semi-flat metric), this data can be read off from the semi-flat metric that we discuss below.

\section{Approximations} \label{sec:approx}

It follows from our previous discussion that, in principle, one can use the complete BPS spectrum of little string theory on $T^3$
to determine smooth K3 metrics.  While such BPS data is `morally attainable,' we do not know the full spectrum at this point in time. 
Therefore, and also based on physical common sense, approximations to the metric given by knowledge of a finite set of BPS charges and masses (say, for a lightest set of BPS states at some point in
moduli space) are a natural thing to discuss.
We begin by reviewing the existing approximations to the desired metric in the large complex structure locus. We then discuss a new approximation and its relationships with the other ones.\footnote{We will not discuss numerical techniques for approximating K3 and more general Calabi-Yau metrics. Interesting work in this direction appears in \cite{headrick:numericK3, donaldson:numericK3}.}

\subsection{Semi-flat metric} \label{sec:semiFlat}

Take an elliptically fibered K3 surface
\be y^2 = x^3 + f(u,w) x + g(u,w) \ , \label{eq:fibration} \ee
where the complex structure of a fiber is specified by
\be j(\tau) := \frac{32 \, \parens{\theta_2^8(\tau) + \theta_3^8(\tau) + \theta_4^8(\tau)}^3}{\theta_2^8(\tau)\theta_3^8(\tau)\theta_4^8(\tau)} = \frac{1}{2} \frac{(24 f)^3}{4f^3 + 27g^2} \ ,\ee
and which has 24 singular fibers (counted with multiplicity) at points $\B^{\rm sing}$ where the discriminant
\be \Delta = 4f^3 + 27g^2 \label{eq:discriminant} \ee
vanishes. Here, $f$ and $g$ are homogeneous polynomials with respective degrees 8 and 12, and so \eqref{eq:fibration} defines a K3 surface via a polynomial of degree 12 in $\WP^3_{1,1,4,6}(u,w,x,y)$. The coordinates $[u,w]$ are the projective coordinates on the base $\B=\PP^1$ of the fibration; we henceforth set $w=1$ (ignoring the point at infinity) and label the positions of the singular fibers as $u_a$. Note that these 24 complex parameters are not independent, as after accounting for both $PSL(2,\CC)$ reparametrizations and scaling $f\to \lambda^2 f$ and $g\to \lambda^3 g$, $f$ and $g$ are specified by 18 complex parameters.

\medskip
The K\"ahler metric of \cite{greene:cosmicString} on this surface is
\be ds^2 = R e^{\phi(u,\bar u)} du d\bar u + \frac{1}{R} \partial \bar\partial K(u,\bar u,z,\bar z) \ ,\ee
where $z\sim z+1\sim z+\tau$ is an elliptic coordinate on the fiber\footnote{This is slightly imprecise, as the fact that $\theta$ is a twisted unitary character will, in general, mean that monodromies can translate $z=0$ to $z=\half$, $z=\frac{\tau}{2}$, or $z=\frac{1+\tau}{2}$. So, really we should call the elliptic coordinate $\tilde z$, relate it to the Wilson-`t Hooft line moduli via $\tilde z=\frac{\tilde\theta_m-\tau \tilde\theta_e}{2\pi}$ (cf. \eqref{eq:zTheta}), and relate it to $z$ via a choice of quadratic refinement, as in \eqref{eq:thetaRel}. We emphasize that locally we may always choose a quadratic refinement (such as the one in \eqref{eq:canonicalRefine}) so that $\tilde z=z$, and so this is only a global issue. In any case, the semi-flat hyper-K\"ahler structure looks the same when expressed in terms of $\theta$ or $\tilde\theta$, since $d\theta_\gamma=d\tilde\theta_\gamma$, and so this distinction is not so important in this section.} and
\be e^\phi = \tau_2 \abs{\eta^2 \prod_{a=1}^{24} (u-u_a)^{-1/12}}^2 \ ,\quad K = -(z-\bar z)^2 / 2\tau_2 \ .\ee
We have introduced the notation $\tau = \tau_1 + i \tau_2$. In particular, $e^\phi$ is modular invariant. More explicitly,\footnote{In this expression, we have corrected a minor error in \cite{greene:cosmicString}.}
\be ds^2 = \brackets{R e^\phi - \frac{(z-\bar z)^2 \partial\tau \bar\partial\bar\tau}{4R\tau_2^3} } du d\bar u + \frac{dz d\bar z}{R\tau_2} - \brackets{ \frac{(z-\bar z)\bar\partial \bar\tau}{2iR\tau_2^2} dz d\bar u + \hdc } \ . \ee
This is called semi-flat because it is flat when restricted to a torus fiber. We also have
\be \partial \bar\partial \phi = \frac{\partial \tau \bar\partial \bar\tau}{(\tau-\bar\tau)^2} = \partial \bar\partial \log \tau_2 \ ,\ee
from which it easily follows that the metric is Ricci-flat (but singular at the points $u_a$):
\be \partial \bar\partial \log \det g_{A \overline{B}} = \partial \bar\partial (\phi - \log \tau_2) = 0 \ .\ee
It has volume form $\omega_+\wedge\bar\omega_+$, where the holomorphic 2-form $\omega_+$ is
\be \omega_+ = \eta^2 \brackets{\prod_{a=1}^{24} (u-u_a)^{-1/12}} du\wedge dz \ .\ee
The special cases with $\tau$ constant are discussed in \cite{sen:FOrientifolds,dasgupta:constant}.

\medskip
We now express the metric and holomorphic 2-form using slightly different notation. First, we introduce locally-defined coordinates $a(u),a_D(u)$, which are periods of the Seiberg-Witten differential. Similarly to \cite{sen:BPS}, we define
\be da = \eta^2 \brackets{ \prod_{a=1}^{24} (u-u_a)^{-1/12} } du \ ,\quad da_D = \tau da \ . \ee
We also introduce $2\pi$-periodic coordinates $\theta_e,\theta_m$:
\be z = \frac{\theta_m - \tau \theta_e}{2\pi} \ . \label{eq:zTheta} \ee
Finally, we introduce the following fiberwise differential that treats $u$ as constant:
\be \tilde dz = \frac{d\theta_m - \tau d\theta_e}{2\pi} = dz + \frac{\theta_e}{2\pi} \partial \tau \,du = dz - \frac{z-\bar z}{2i\tau_2} \partial \tau \,du \ .\ee
Then, we have
\be \omega_+ = da \wedge dz = da \wedge \tilde dz \ee
and
\be ds^2 = R \tau_2 \, da d\bar a + \frac{\tilde dz \tilde d\bar z}{R\tau_2} \ .\ee
The latter corresponds to a K\"ahler form
\be \omega_3 = \frac{i}{2}\parens{ R \tau_2 \, da \wedge d \bar a + \frac{1}{R\tau_2} \, \tilde dz\wedge \tilde d\bar z} \ .\ee
These expressions are precisely those in \cite{GMN:walls} obtained by dimensionally reducing the 4d IR abelian gauge theory on a circle (ignoring KK modes) and dualizing the 3d gauge field to a scalar, $\theta_m$.

\medskip
We now choose a symplectic basis $\{\gamma^m,\gamma^e\}$ with $\avg{\gamma^m,\gamma^e}=1$, and introduce the relabeling $Z_{\gamma^e} = a$, $Z_{\gamma^m} = a_D$. Then, one has
\be \omega_+ = - \frac{1}{2\pi} \epsilon_{ij} \, dZ_{\gamma^i} \wedge d\theta_{\gamma^j} \ee
and
\be \omega_3 = \frac{R}{4}\epsilon_{ij}\, dZ_{\gamma^i}\wedge d\overline{Z_{\gamma^j}} - \frac{1}{8\pi^2R} \epsilon_{ij}\, d\theta_{\gamma^i}\wedge d\theta_{\gamma^j} \ .\ee
These are the results one obtains from using $\X_\gamma^{\rm sf}$ to obtain the holomorphic symplectic form $\varpi(\zeta)$ \cite{GMN:walls}. The mass parameters $Z_{\gamma^f}$ (where $\gamma^f\in\Gamma_{\rm flavor}$) can similarly be extracted from the above formulae by considering the values of $a$ and $a_D$ at a singular fiber -- or, more specifically, by considering the linear combination thereof given by the gauge charges of a BPS state that becomes massless at the singular fiber.

\medskip
Finally, we note that the volumes of the non-singular fibers are $1/R$. This concretely illustrates the geometric meaning of the large complex structure limit.

\subsection{The Ooguri-Vafa and Gross-Wilson metrics}

We now restrict to a generic configuration of 7-branes, where no two $u_a$ coincide. This is stronger than requiring that the total space be smooth, as in this case one can have either $II$ or $I_1$ fibers, whereas we are only allowing 24 $I_1$ fibers. (Physically, $II$ fibers correspond to an H7-plane, which is formed by merging mutually non-local 7-branes. The worldvolume field theory of a D3-brane probe of this configuration is described in \cite{sw:argyresDouglas}.) The reason for this restriction is that now, near each 7-brane there is a choice of electromagnetic duality frame such that the physics is locally $U(1)$ with a charge 1 hypermultiplet.

\medskip
Since the latter has only electrically charged BPS states, the instanton corrections to the moduli space $\M$ on $\RR^3\times S^1_R$ can be straightforwardly computed \cite{vafa:spacetimeInsts,seiberg:mirrorT}. The metric one obtains on $\M$ is called the Ooguri-Vafa metric. It is smooth even at $\M\backslash \M'$, reflecting the absence of a Higgs branch in this gauge theory. Furthermore, at large $R$ it decays very rapidly away from $\B^{\rm sing}$ to a semi-flat metric.

\medskip
Of course, the theory is only an effective field theory; this manifests itself via the fact that $\B$ has a cutoff $\Lambda$:
\be \B = \{ u\in \CC : |u| < |\Lambda| \} \ , \quad \B' = \B \backslash \{0\} \ .\ee
We choose a symplectic basis $\{\gamma^m,\gamma^e\}$ for $\Gamma_u$ (with $\avg{\gamma^m,\gamma^e}=1$) so that $\gamma^e$ has trivial monodromy and $\gamma^m\to \gamma^m + \gamma^e$ around $u=0$. $\Gamma_{\rm flavor}$ is trivial. The BPS index is simply $\Omega(\pm \gamma^e; u) = 1$ for all $u\in\B'$; $\Omega$ vanishes for all other charges. The central charges are $Z_{\gamma^e}=u$ and $Z_{\gamma^m} = \frac{u}{2\pi i}(\log (u/\Lambda) - 1)$.

\medskip
Because all BPS states are mutually local, the approximation \eqref{eq:firstApprox} is exact, and furthermore the final term vanishes. Also, $\X_{\gamma^e} = \X^{\rm sf}_{\gamma^e}$. We therefore have
\be \varpi(\zeta) = \varpi^{\rm sf}(\zeta) + \sum_{\gamma = \pm \gamma^e} \varpi^{\rm inst}_\gamma(\zeta) \ ,\ee
where $\varpi^{\rm inst}_\gamma(\zeta)$ is given by \cite{GMN:walls}
\begin{align}
\varpi^{\rm inst}_\gamma(\zeta) &= \frac{1}{16\pi^3 i R}\epsilon_{ij} d'\Y^{{\rm sf, \, }i}(\zeta) \wedge \avg{\gamma^j, \gamma} \int_{\ell_\gamma(u)} \frac{d\zeta'}{\zeta'}\frac{\zeta'+\zeta}{\zeta'-\zeta} \,\frac{d'\X^{\rm sf}_\gamma(\zeta')}{1-\X^{\rm sf}_{\gamma}(\zeta')} \\
&= - \frac{i}{8\pi^2} d'\Y^{{\rm sf}}_\gamma(\zeta) \wedge \brackets{ - A^{\rm inst} d\log\parens{Z_\gamma/\overline{Z_\gamma}} + V^{\rm inst}\parens{\frac{1}{\zeta} dZ_\gamma - \zeta d\overline{Z_\gamma}}} \ , \label{eq:wInst} \\
A^{\rm inst} &= \sum_{n>0} e^{in\theta_\gamma} |Z_\gamma| K_1(2\pi R n |Z_\gamma|) \ ,\\
V^{\rm inst} &= \sum_{n>0} e^{in\theta_\gamma} K_0(2\pi R n |Z_\gamma|) \ .
\end{align}
This determines the Ooguri-Vafa metric. The asymptotics $K_\nu(x) \sim \sqrt{\frac{\pi}{2x}} e^{-x}$ as $x\to\infty$ show that instanton corrections are exponentially suppressed away from the origin ($u=0$), as expected.

\medskip
With this in hand, we can explain the approximation of Gross and Wilson \cite{gross:OV} to a smooth Ricci-flat K3 metric. The idea is to correct the semi-flat metric by gluing in Ooguri-Vafa metrics at the singular fibers to obtain a result that is globally defined, but not quite Ricci-flat. Somewhat more precisely, they show that for sufficiently large $R$, one can construct a K\"ahler metric on $\M$ (which they take to be a genus one fibration) with the following properties. First, there exist collections $\{U_1^a\}_{a=1}^{24}$ and $\{U_2^a\}_{a=1}^{24}$ of open sets in $\B=\PP^1$, where $U_1^a$ and $U_2^a$ enclose (only) the $a$-th singular point and $U_1^a\subset U_2^a$, such that over each $U_1^a$ one has the Ooguri-Vafa metric and on the complement of $\cup_a U_2^a$ one has a semi-flat metric. Second, each non-singular fiber has volume $1/R$. And third, the metric is (in a precise sense) exponentially close to Ricci-flat. In particular, Gross and Wilson show that if one starts with their approximate metric and improves it to the Ricci-flat one whose K\"ahler class is in the same cohomology class (following Yau's proof \cite{yau:CY} of the Calabi conjecture), then the difference between these metrics is quite small away from the singular fibers.\footnote{This reasoning suggests the following amusing approach to determining the BPS invariants. For a number of complex structures, begin with the Gross-Wilson metric and numerically solve the complex Monge-Amp\`ere equation in order to approximate the K\"ahler form. One might then be able to extract the $\X_\gamma$'s and the BPS invariants, at least for small charges.} (See also Corollary 4.32 and Remark 4.33 of \cite{lin:walls2}.)

\subsection{New nearly Ricci-flat and semi-flat metrics} \label{sec:pGW}

We begin by noting that even the approximation \eqref{eq:firstApprox} contains far more contributions than are necessary in order to find a sensible metric at large $R$. First, the final term is clearly subleading compared to the second term, as the former is second order in corrections to the semi-flat metric while the latter is only first order in such corrections. Second, even the first order correction term is excessive, as almost all 4d BPS states remain massive everywhere in $\B$ and so their contributions are everywhere exponentially suppressed at large $R$. Only 48 states ever really matter\footnote{We're looking at you, Indiana and New Jersey.} -- namely, the 2 states that contribute to the Ooguri-Vafa metric that is almost exactly valid near each of the 24 singular fibers. We denote the set of (gauge and flavor) charges of such states by $\hat\Gamma_u^0$. Finally, we use the fact that near the singular fiber where one of these states is light, the corresponding charge has $\Omega=1$; away from the singular fiber, the state is irrelevant, and so we can set $\Omega=1$ everywhere.

\medskip
We then obtain the following `pseudo-Gross-Wilson' approximation to the holomorphic symplectic form:
\be \varpi^{\rm pGW}(\zeta) = \varpi^{\rm sf}(\zeta) + \frac{1}{16\pi^3 i R}\epsilon_{ij} d'\Y^{{\rm sf, \, }i}(\zeta) \wedge \sum_{\gamma \in \hat\Gamma^0_u} \avg{\gamma^j, \gamma} \int_{\ell_\gamma(u)} \frac{d\zeta'}{\zeta'}\frac{\zeta'+\zeta}{\zeta'-\zeta} \,\frac{d'\X^{\rm sf}_\gamma(\zeta')}{1-\X^{\rm sf}_{\gamma}(\zeta')} \ . \label{eq:pGW} \ee
Following \S5.6 of \cite{GMN:walls}, we exploit the fact that our charges are in $\hat\Gamma^0_u$ by choosing, for each $\gamma$, a symplectic basis $\{\gamma^i\}$ for $\Gamma_u$ such that $\gamma \in \gamma^1\oplus \Gamma_{\rm flavor}$. (If instead our charges were in $\hat\Gamma_u$, we could only choose $\gamma\in (q^\gamma \gamma^1)\oplus \Gamma_{\rm flavor}$ for some $q^\gamma\in \ZZ$.) Then, \eqref{eq:pGW} reduces to
\be \varpi^{\rm pGW}(\zeta) = \varpi^{\rm sf}(\zeta) + \sum_{\gamma\in \hat\Gamma_u^0} \varpi_\gamma^{\rm inst}(\zeta) \ ,\ee
where $\varpi_\gamma^{\rm inst}$ is as in \eqref{eq:wInst}.

\medskip
Next, we use \eqref{eq:metric} (keeping only the leading corrections to the semi-flat metric) to approximate the metric on $\M'$. This new metric is presumably in some ways inelegant -- e.g., it is unlikely to be K\"ahler. Nevertheless, it has a number of desirable properties. Namely, at large $R$ this simple approximation -- which requires no knowledge of the BPS invariants -- is almost identical to the Ricci-flat hyper-K\"ahler metric on $\M'$ which is nearly semi-flat away from singular fibers and nearly Ooguri-Vafa near singular fibers.

\medskip
This makes it clear that our approximation is quite similar to that of Gross and Wilson. The most important difference, as we see it, is that there is a way to systematically improve our approximation to find the Calabi-Yau metric on $\M'$.

\medskip
We note that although we restricted here, for simplicity (and because this is the generic configuration), to semi-flat limits with 24 $I_1$ fibers, there is an analogous approximation that applies in the presence of arbitrary collections of $I_N$ singular fibers (where $N$ can differ between the singular fibers). One simply inserts knowledge of the $N$ hypermultiplets which become massless as the D3-brane approaches the singular fibers. These are present in the spectra of the associated $U(1)$ field theories with $N$ charge 1 hypermultiplets describing D3-branes probing these singular fibers. For more general singular fibers, the metric should still locally be approximated by that associated to the D3-brane probe field theory, but now due to mutually non-local BPS states the determination of the field theory metric requires solving a non-trivial integral equation.

\section{Conclusion} \label{sec:conclude}

In this work, we have exploited many of the known properties of little string theories in order to reduce the long-standing problem of determining a Ricci-flat metric for a non-toroidal compact Calabi-Yau manifold to the computation of a BPS spectrum. Our approach combines many of the ideas that have been developed for studying the large complex structure limit of Calabi-Yau manifolds \cite{strominger:mirrorT,gross:OV,fukaya:insts,KS:gluing,GrossSiebert} with the observation of Gaiotto, Moore, and Neitzke that the twistor space associated to a hyper-K\"ahler manifold provides a powerful means for obtaining a metric \cite{GMN:walls}.

\medskip
Two general lessons suggest themselves. First, the idea that moduli spaces of supersymmetric theories can be usefully studied in terms of the rings of functions on them is certainly not new \cite{ADS,ADS2,taylor:varieties}, but the present application suggests that physics can benefit from algebro-geometric approaches to building up manifolds as ringed spaces. Indeed, this has been exploited recently in the study of 3d Coulomb branches \cite{hanany:coulombHS,gaiotto:coulomb,nakajima:coulomb1,nakajima:coulomb2}. Secondly, the constraints of wall crossing formulae are extremely powerful and can be used to derive a number of non-trivial results. This too has been exploited recently in another context \cite{mz:walls}.

\medskip
We conclude with suggestions for future work.

\begin{itemize}
\item While it is physically clear that our metric can be extended over the singular fibers, since the moduli space of the theory on $S^1_R$ is generically a smooth K3 surface, we do not demonstrate this explicitly. This should be achievable by generalizing the arguments of \cite{garza:singular1,garza:singular2}.

\item The other missing ingredient in the determination of the metric is the computation of the invariants $\Omega(\gamma;u)$. In principle, the Gromov-Witten approach of \cite{lin:walls1,lin:walls2,lin:walls3,lin:walls4,lin:walls5} ought to suffice. However, the reformulation of these invariants as BPS state counts in a little string theory suggests additional means by which they might be determined. For example, indices of compactified little string theories have been studied from a holographic approach in \cite{Harvey:2013mda,Harvey:2014cva,harvey:lstThermo}, and heterotic little string theories in particular have been studied holographically in \cite{kapustin:hlst,narain:hlst}. (However, this has the drawback that the holographic approach misses non-perturbative states \cite{s:ns5holo,kapustin:hlst}.) A DLCQ formulation also exists \cite{kachru:dlcq,lowe:dlcq}. The computations of \cite{kim:HLST} might also be generalizable.

The approach we suspect will be the most useful is geometrically engineering the little string theory with F-theory on a non-compact Calabi-Yau 3-fold, as in \cite{vafa:F2,aspinwall:hetF,sk:tate,vafa:fClass}. After compactifying the little string theory on $T^2$, one finds a type IIA description (and a mirror IIB description), as was recently exploited for field theories in \cite{vafa:6d4d}. T-duality might provide a useful means for studying the theory in the $R\to 0$ limit. Geometric engineering might provide a useful (e.g., quiver) characterization of BPS states, as it does for field theories \cite{fiol:fractBrane,gukov:refineMotive2,Chuang:2013wt}. It might also yield a `spectral curve' for K3 surfaces, in the sense of footnote \ref{ft:spectral}.

\item Continuing this last point, we observe that the characterization of BPS states of a 4d field theory via geodesics on its Seiberg-Witten curve \cite{vafa:geodesics,brand:geodesics,GMN:classS,GMN:networks,GMN:snakes} provides an extremely powerful formalism for determining the metric for all $R$ \cite{GMN:classS}. It would be fantastic if this set of ideas generalized to the study of compactified little string theories and their K3 moduli spaces.

\item There is an interesting proof of the wall crossing formula in \cite{vafa:topWalls} where the twistor space appears in spacetime. This work also relates 4d $\N=2$ class S BPS state counting problems to the determination of a superpotential via open topological string theory. Implicit in \cite{lin:thesis,lin:walls4} is the observation that this generalizes to our little string theory. It is also explained how to make this approach to BPS state counting rigorous, using real Noether-Lefschetz theory. This could conceivably give another route toward solving our counting problem.

\item There are many little string theories besides the heterotic 5-brane considered in this paper \cite{vafa:fClass}. 
For instance, \cite{intriligator:compact} discusses variants whose moduli spaces coincide with the moduli spaces of instantons on K3
or $T^4$.
It might be interesting to employ the approach of this paper to study them as well. Lastly, we note that the results of \cite{ganor:noncomm1,ganor:noncomm2} imply that many K3 metrics can be determined from BPS state counts (flavored by an R-symmetry which is a flavor symmetry from the point of view of 4d $\N=2$ supersymmetry, i.e. the index $B_2(z)$ of \cite{sen:N4refine}) in a $T^2$-compactified $\N=(1,1)$ little string theory.

\item Most ambitiously, one might hope that this set of ideas can be generalized to say something about non-hyper-K\"ahler Calabi-Yau threefolds near their large complex structure point. Field theory probes thereof have been studied \cite{douglas:curvedMatrix,sk:curvedMatrix}; can they be little string-ified? If so, one might hope that little string theory could be of use to the Gross-Siebert program.


\end{itemize}

\section*{Acknowledgments}

We thank G. Moore and A. Neitzke for comments on our draft; we understand that they, along with D. Gaiotto, had informal discussions about
similar directions in the past.
A.T. gratefully acknowledges many useful talks by and conversations with Yu-Shen Lin on his work, as well as many insightful comments by Shing-Tung Yau, Yu-Wei Fan, and the rest of the Yau group at Harvard/CMSA on related ideas. 
The research of S.K. was supported in part by a Simons Investigator Award and the National Science Foundation under grant number PHY-1720397.  The research of A.T. was supported by the National Science Foundation under NSF MSPRF grant number 
1705008.

\newpage
\appendix

\bibliography{Refs}

\end{document}